\documentclass[12pt]{article}
\usepackage{amsmath}
\usepackage{epsfig,float}
\usepackage{amsmath}
\usepackage{amssymb}
\usepackage[all]{xy}
\usepackage{verbatim}
\usepackage{epsfig,amsmath,amsfonts,amssymb,amsthm,amstext,amscd,eucal}

\begin{document}

\title{Heun functions describing bosons and fermions on Melvin's spacetime}

\author{Marina--Aura Dariescu and Ciprian Dariescu \\
Faculty of Physics,
^^ ^^ Alexandru Ioan Cuza'' University of Ia\c{s}i \\
Bd. Carol I, no. 11, 700506 Ia\c{s}i, Romania}

\date{}
\maketitle

\begin{abstract}
Employing a pseudo-orthonormal coordinate-free approach, the solutions to the Klein--Gordon and Dirac equations for particles in Melvin spacetime are derived in terms of Heun's biconfluent functions.
\end{abstract}

\begin{flushleft}
\end{flushleft}

\baselineskip 1.5em

\newpage

\section{Introduction}

The study of relativistic particles in static magnetic fields has a long history and is still attracting considerable attention, especially for cases where someone deals with curved manifolds.

Even though on Minkowski spacetime the 
relativistic behavior of an electron in various magnetostatic configurations
is well understood (see, for example, the Johnson and Lippmann's paper [1]),
a weakness on curved spacetime regards the explicit gauge-covariant formulation.  

Recently, when dealing with slowly rotating neutron stars which have been termed as magnetars [2], it has been assumed that their huge magnetic induction in the core and crust, $B \sim 10^{14} - 10^{15}$ (G), is affecting the spacetime geometry. A way out could be the search for general relativistic solutions with the magnetic field considered as a perturbation of the spherically symmetric background [3].
Another way is to assume that magnetized metrics, as the one belonging to the Melvin class [4, 5] may be reliable candidates for describing these
highly compact astrophysical objects with a dominant axial magnetic field
[6].

Within a coordinate-dependent formulation, switching between canonical and pseudo-orthonormal basis, the above mentioned authors are
integrating the system of four coupled first-order differential equations, in the first approximation, neglecting the terms in higher orders of the polar radial coordinate $\rho$.
Their solutions are expressed in terms of generalized Laguerre polynomials, similarly to the case of the Dirac equation in cylindrical coordinates on a flat manifold  [7].

In the present work, we are applying a free of coordinates method to  analyze the Klein--Gordon and Dirac equations describing particles evolving in a Melvin's spacetime.
Employing the Cartan's formalism, we are computing all the essential geometrical objects for writing down the corresponding matter fields and Einstein's equations.

It turns out that the $SO(3,1) \times U(1)-$gauge covariant Klein--Gordon equation can be exactly solved, its solutions being given by the Heun biconfluent functions 
[8--10]. The same happens with the approximate expression of the second-order differential system derived from the Dirac equation.

The Heun functions, either general or confluent, are main targets of recent investigations and have been obtained for massless particles evolving in an Universe described by the metric function written as a non-linear mixture of Schwarzschild, Melvine and Bertotti-Robinson solutions  [11].

\section{The geometry}

Recently, in [12], it was presented the procedure of transforming a
known static symmetric solution to Einstein-hydrodynamic equations  into a magnetized
metric, by (nonlinearly) adding the magnetic field. In spherical coordinates, this has the general form
\begin{equation}
ds^2= \Lambda^2 g_{11}  (dr)^2 + \Lambda^2 r^2  (d \theta )^2 + \frac{r^2 \sin^2 \theta}{\Lambda^2}  (d \varphi)^2 - \Lambda^2 g_{00}  (dt)^2 \, ,
\end{equation}
with the metric functions $g_{11}$ and $g_{00}$ depending only on $r$ and 
\begin{equation}
\Lambda = 1 + \frac{B_*^2}{4} r^2 \sin^2 \theta \,  ,
\end{equation}
where, for the moment, $B_*$ is a parameter related to the magnetic field intensity.

In the pseudo-orthonormal Cartan frame corresponding to the metric (1),
\begin{equation}
\omega^1=  \Lambda \sqrt{g_{11}} \, dr\; , \; \;
\omega^2= \Lambda r  \,d\theta \; , 
\; \;
\omega^3=\frac{r \sin \theta}{\Lambda} \, d\varphi \; , \; \;
\omega^4=\Lambda \sqrt{g_{00}} \,dt \; ,
\end{equation}
for the potential 
\begin{equation}
A_2= \frac{B_0 r \sin \theta}{2} \, ,
\end{equation}
where $B_0$ is the strength of the magnetic field on the axis,
the Maxwell tensor components, corresponding to a poloidal magnetic field with $B_{\rho}$ and $B_{\theta}$, are given by the relations 
\begin{equation}
F_{23} = \frac{B_0 \cos \theta}{\Lambda^2} \; , \;
F_{13} = \frac{B_0 \sin \theta}{\Lambda^2 \sqrt{g_{11}}} \; ,
\end{equation}
pointing out a prolate (in shape) star.

Once we assume $g_{00} = g_{11} =1$, we can switch to cylindric coordinates $\lbrace \rho , \varphi , z , t \rbrace$, by
\[ 
\rho = r \sin \theta \; , \;  \;  z = r \cos \theta \, ,
\]
so that the magnetized metric (1) turns into the simple Melvin expression 
\begin{equation}
ds^2 \, = \, \Lambda^2  ( d \rho )^2 + \frac{\rho^2}{\Lambda^2} ( d \varphi)^2 + \Lambda^2 (dz)^2  - \Lambda^2 (dt)^2 \, ,
\end{equation}
with
\begin{equation}
\Lambda = 1 + \frac{B_*^2 \rho^2}{4}  \; .
\end{equation}

Within an $SO(3,1)-$gauge covariant formulation, we introduce the pseudo-orthonormal frame
\[
e_1 = \frac{1}{\Lambda} \partial_{\rho} \; , \; e_2 = \frac{\Lambda}{\rho} \partial_{\varphi} \; , \; e_3 = \frac{1}{\Lambda} \partial_z \; , \;
e_4 = \frac{1}{\Lambda} \partial_z \; , \]
whose corresponding dual base is 
\begin{equation}
\omega^1=  \Lambda \, d \rho \; , \; \;
\omega^2 = \frac{\rho}{\Lambda} \, d\varphi \; , \; \;
\omega^3 = \Lambda \,dz \; , 
\; \;
\omega^4= \Lambda \,dt \;  ,
\end{equation}
so that the metric (6) gets the Minkowskian form $ds^2 = \eta_{ab} \omega^a \omega^b$, with $\eta_{ab} = \left[ 1 , \, 1 , \, 1 , \, -1 \right]$.
The first Cartan equation, 
\begin{eqnarray}
& &
d\omega^a=\Gamma^a_{.[bc]}\,\omega^b\wedge \omega^c \, ,
\end{eqnarray}
with $1 \leq b<c \leq4$ and $\Gamma^a_{.[bc]} = \Gamma^a_{.bc} - \Gamma^a_{. cb}$,
leads to the following connection one-forms:
\begin{equation}
\Gamma_{12} = \left( \frac{\Lambda^{\prime}}{\Lambda^2} - \frac{1}{\rho \Lambda} \right) \,\omega^2 \; , \; \;
\Gamma_{13} = - \, \frac{\Lambda^{\prime}}{\Lambda^2} \, \omega^3 \; , \; \;
\Gamma_{14} = \frac{\Lambda^{\prime}}{\Lambda^2} \, \omega^4 \; , \;
\end{equation}
where $\Lambda^{\prime}$ is the derivative of $\Lambda$ with respect to $\rho$.

Employing the second Cartan equation
\begin{eqnarray}
& &
\mathcal{R}_{ab}=d\Gamma_{ab}+\Gamma_{ac}\wedge\Gamma^c_{.b} ,
\end{eqnarray}
one derives the curvature two-forms $\mathcal{R}_{ab} = R_{abcd} \, \omega^c \wedge \omega^d$, with $1 \leq c<d \leq 4$, leading to the curvature components 
\begin{eqnarray}
& &
R_{1212} = \frac{2B_*^2}{\Lambda^4} \left[ 1 - \frac{B_*^2 \rho^2}{8} \right] , \; \;
R_{3434} = \frac{B_*^4 \rho^2}{4 \Lambda^4} 
\; , \nonumber \\*
& &
R_{1313} = R_{2323} = - \, \frac{B_*^2}{2 \Lambda^4} \left[ 1 - \frac{B_*^2 \rho^2}{4} \right]  = - R_{1414} = - R_{2424} \, , \nonumber
\end{eqnarray}
pointing out the special radius value $\rho_* = 2/B_*$, for which only the components $R_{1212} = B_*^2 / \Lambda^4 = R_{3434} $ are surviving and the Weyl tensor vanishes.

Since the scalar curvature is zero, the Einstein tensor components
are given by the Ricci tensor components, as
\begin{equation}
G_{11} = G_{22} = - \, G_{33} = G_{44} = \frac{B^2_*}{\Lambda^4} \, .
\end{equation}

In the pseudo-orthonormal frame whose dual bases is (8), it turns out that the potential (4), generating the magnetic induction along $Oz$, gets the familiar expression
\begin{equation}
A_2 = \frac{B_0 \rho}{2}  \,  ,
\end{equation} 
and the essential component of the Maxwell tensor reads
\[
F_{12} = A_{2|1} + \Gamma_{212} A_2 = \frac{B_0}{\Lambda^2} \, ,
\]
where $f_{|1} = e_1 (f)$.

Using the energy-momentum tensor components
\[
T_{11} = T_{22} = - \, T_{33} = T_{44} = \frac{1}{2} F_{12}^2 =
\frac{1}{2} \frac{B_0^2}{\Lambda^4} \, ,
\]
in the Einstein equations $G_{ab} = \kappa_0 T_{ab}$,
one gets the following relation between the parameters $B_*$ and $B_0$, 
\begin{equation}
B_*^2 = \frac{\kappa_0 B_0^2}{2} \, ,
\end{equation}
with $\kappa_0 = 8 \pi G /c^4$.

\section{Exactly solvable Klein--Gordon equation}

In this section, we are going to construct the wave function of the charged bosons, considered as test particles evolving in the crust of a relativistic magnetar.
The complex scalar field of mass $\mu$,
minimally coupled to gravity, is described by the $SO(3,1) \times U(1)$ gauge-covariant Klein--Gordon equation
\[
\eta^{ab} \Phi_{|ab} - \eta^{ab} \Phi_{|c} \Gamma^c_{\; ab} \, = \, \mu^2 \Phi \,  + 2iq \, A_2 \, \Phi_{|2} + q^2 A_2^2 \, \Phi \, ,
\]
which, in the
pseudo-orthonormal frame with the dual bases (8), reads
\begin{eqnarray}
& &
\frac{1}{\rho} \frac{\partial \;}{\partial \rho} \left[ \rho \frac{\partial \Phi}{\partial \rho} \right]
+ \frac{\Lambda^4}{\rho^2} \frac{\partial^2 \Phi}{\partial \varphi^2} 
+ \frac{\partial^2 \Phi}{\partial z^2} 
- \frac{\partial^2 \Phi}{\partial t^2} 
\nonumber \\* & & = \left[
\mu^2 \Lambda^2 + 
iq \, B_0 \Lambda^3 \frac{\partial \Phi}{\partial \varphi} + \left( \frac{q B_0 \rho \Lambda }{2} \right)^2 \right] \Phi  \, .
\end{eqnarray}
The above form suggests the variables separation
\begin{equation}
\Phi = \phi (\rho) e^{im \varphi} e^{i p_z z} e^{-i \omega t} \, ,
\end{equation}
which leads to the following differential equation for the un-known function $\phi$,
\begin{equation}
\frac{1}{\rho} \frac{\partial \;}{\partial \rho} \left[ \rho \frac{\partial \phi}{\partial \rho} \right] + \left[ \omega^2 -  p_z^2 - \frac{m^2}{\rho^2} \Lambda^4 - \mu^2 \Lambda^2 + m qB_0 \Lambda^3  - \left( \frac{qB_0 \rho \Lambda}{2} \right)^2  \right] \phi = 0 \, ,
\end{equation}
with $\Lambda$ defined in (7).

This can be exactly integrated, its solution being expressed in terms of the Heun biconfluent function as [9, 10]
\begin{equation}
\phi (y) \sim \exp \left[ - \frac{y^2}{2}  - \frac{\beta y}{2}  \right] y^{\alpha /2} HeunB \left[ \alpha , \, \beta , \, \gamma , \, \delta , \, y \right] ,
\end{equation}
where the variable and the parameters are respectively given by
\begin{equation}
y = \frac{\sqrt{b} \, B_* \rho^2}{4} \; , \; \; b = q B_0 - \frac{mB_*^2}{2} \approx q B_0 \, ,
\end{equation}
and
\begin{equation}
\alpha = \pm m \; , \;
\beta \approx 2 \frac{\sqrt{qB_0}}{B_*}  \; ,
\; 
\gamma \approx m - \frac{\mu^2}{qB_0} \; ,
\; 
\delta \approx - \, \frac{2 \left[\omega^2 - p_z^2 - \mu^2 + m q B_0 \right]}{B_* \sqrt{q B_0}} \;  .
\end{equation}

Let us point out that the Heun biconfluent equation has one regular singularity at the origin and one irregular at $\infty$ and can be obtained, from the Heun general equation, by a process of successive confluences [10].

Regarding the asymptotic behavior of the function (18), solution to the equation (17), that has a singularity in $\rho \to 0$, due to the exponential term, this is vanishing for large $y-$values.  On the other hand, for a regular solution at the origin $y=0$ (where $HeunB(0) =1$), one has to choose the plus sign of $\alpha$ in (20).

\section{The $SO(3,1) \times U(1)-$gauge covariant Dirac Equation}

For relativistic fermions of mass $M$, coupled to the external magnetic field generated by (13), the Dirac equation has the $SO(3,1) \times U(1)-$gauge covariant expression
\begin{equation}
\gamma^a \, \Psi_{;a} + M \Psi \, = 0 \, ,
\end{equation}
where ^^ ^^ ;'' stands for the covariant derivative
\begin{equation}
\Psi_{;a} =  e_a \Psi + \frac{1}{4} \, \Gamma_{bca} \, \gamma^b  \gamma^c \Psi - i q A_a \Psi \; .
\end{equation}
In view of the relations (10), the term expressing the Ricci spin-connection in the equation (21) reads
\[
 \frac{1}{4} \, \Gamma_{bca} \, \gamma^a \gamma^b \gamma^c = \frac{f}{\Lambda}  \,  \gamma^1  \, ,
 \]
where we have introduced the function
\begin{equation}
f = \frac{1}{2} \left[ \frac{1}{\rho} + \frac{\Lambda^{\prime}}{\Lambda} \right] \, .
\end{equation}

The explicit form of the Dirac equation (21) being
\begin{equation}
\frac{1}{\Lambda} \left[
\gamma^1 \left( \partial_{\rho} + f \right) + \frac{\Lambda^2}{\rho} \gamma^2 \partial_{\varphi} + \gamma^3 \partial_z + \gamma^4 \partial_t + M \Lambda - i q \gamma^2 \Lambda A_2 \right] \Psi = 0 \, ,
\end{equation}
one may use the variables separation
\begin{equation}
\Psi = e^{i(m \varphi + p_z z - \omega t)} \psi ( \rho ) \; , 
\end{equation}
to derive the differential equation satisfied by the part depending on $\rho$, i.e.
\begin{equation}
\gamma^1 \left[ \psi^{\prime} + f \psi \right] +  i \left \lbrace  \gamma^2 \Lambda
\left[ \frac{m \Lambda}{\rho} - \frac{qB_0 \rho}{2}  \right] +  p_z \gamma^3 - \omega \gamma^4 - i M \Lambda  \right \rbrace \psi = 0 \, .
\end{equation}

With the following function substitution
\begin{equation}
\psi = \frac{1}{2 \sqrt{\rho \Lambda}} \, \chi \, ,
\end{equation}
the above equation becomes
\begin{equation}
\gamma^1 \chi^{\prime} +  i \left \lbrace  \gamma^2 \, F
+  p_z \gamma^3 - \omega \gamma^4 - i M \Lambda  \right \rbrace \chi = 0 \, ,
\end{equation}
where
\begin{equation}
F ( \rho ) = \Lambda \left[ \frac{m \Lambda}{\rho} - \frac{qB_0 \rho}{2}  \right] ,
\end{equation}
and we are going to use the Dirac representation for the $\gamma^i$ matrices,
\begin{equation}
\gamma^{\mu} = -i \beta \, \alpha^{\mu} \; , \; \; \gamma^4
= - i \beta \; , \; \; \mu = \overline{1,3}
\, ,
\end{equation}
with
\[
\beta = \left( \begin{array}{cc} {\cal I} & 0 \\ 0 & - {\cal I}
\end{array} \right)
\; , \; \; \alpha^{\mu} = \left(
\begin{array}{cc}
0 & \sigma^{\mu} \\
\sigma^{\mu} & 0
\end{array}
\right) ,
\]
where $\sigma^{\mu}$ denotes the usual Pauli matrices.

In the followings, we are assuming that the particle is not moving along the magnetic field direction, i.e. $p_z = 0$, and the bi-spinor $\chi$ is of the form 
\begin{equation}
\chi( \rho ) = \left[ \begin{array}{c}
\zeta  ( \rho ) \\ \eta ( \rho )
\end{array} \right] ,
\end{equation}
so that the equation (28) decouples in two equations for the (two-component) spinors $\zeta$ and $\eta$ i.e.
\begin{eqnarray}
& &
\sigma^1 \zeta^{\prime} + i F \sigma^2 \zeta  = i ( \omega + M \Lambda ) \eta \; ,
\nonumber \\*
& &
\sigma^1 \eta^{\prime} + i F \sigma^2 \eta  = i ( \omega - M \Lambda ) \zeta \; .
\end{eqnarray}

Applying the usual procedure, one gets the following differential equations
\begin{equation}
\zeta_A^{\prime \prime} - \frac{M \Lambda^{\prime}}{\omega + M \Lambda} \, \zeta^{\prime}_A + \left \lbrace \omega^2 - M^2 \Lambda^2 - F^2 \mp \left[ F^{\prime} - \frac{M F \Lambda^{\prime}}{\omega + M \Lambda} \right] \right \rbrace \zeta_A = 0 \; , 
\end{equation}
and
\begin{equation}
\eta_A^{\prime \prime} + \frac{M \Lambda^{\prime}}{\omega - M \Lambda} \, \eta^{\prime}_A + \left \lbrace \omega^2 - M^2 \Lambda - F^2 \mp \left[ F^{\prime} + \frac{M F \Lambda^{\prime}}{\omega - M \Lambda} \right] \right \rbrace \eta_A = 0 \; ,
\end{equation}
which cannot be analytically solved. However,
by imposing the condition $B_*^2 \ll q B_0$ and neglecting the powers of $\rho$ larger than 3, the equations (33) and (34) get the simpler forms
\begin{eqnarray}
& &
\zeta_A^{\prime \prime} - \frac{M B_*^2 \rho}{2 (\omega + M)} \left[ 1 - \frac{M B_*^2 \rho^2}{4( \omega +M)} \right] \zeta^{\prime}_A \nonumber
\\*
& & + \left \lbrace \omega^2 - M^2 + \left( m \pm \frac {1}{2} \right) q B_0 - \frac{m(m \mp 1)}{\rho^2} - \left( \frac{qB_0 \rho}{2} \right)^2  \right \rbrace \zeta_A = 0 \; , 
\end{eqnarray}
and
\begin{eqnarray}
& &
\eta_A^{\prime \prime} + \frac{M B_*^2 \rho}{2 (\omega - M)} \left[ 1 + \frac{M B_*^2 \rho^2}{4( \omega - M)} \right] \eta^{\prime}_A \nonumber
\\*
& & + \left \lbrace \omega^2 - M^2 + \left( m \pm \frac {1}{2} \right) q B_0 - \frac{m(m \mp 1)}{\rho^2} - \left( \frac{qB_0 \rho}{2} \right)^2  \right \rbrace \eta_A = 0 \; , 
\end{eqnarray}
The corresponding solutions, i.e.
\begin{eqnarray}
& &
\zeta_1 = \left \lbrace \sqrt{\rho} , \rho^{m} \right \rbrace
u_1 \; , \; \;
\zeta_2 = \left \lbrace \sqrt{\rho} ,  \rho^{m+1} \right \rbrace u_2 \; , \; \nonumber \\*
& &
\eta_1 = \left \lbrace \sqrt{\rho} ,  \rho^m \right \rbrace v_1 \; , \; \eta_2 = \left \lbrace \sqrt{\rho} ,  \rho^{m+1} \right \rbrace v_2 \, ,
\end{eqnarray} 
are
expressed in terms of Heun's biconfluent functions [8--10]
\begin{eqnarray}
& &
u_1 = HeunB \left[ \alpha_1 , \beta , \gamma^+ , \delta^+_1 ,  \, x_u \right] \; , \;
u_2 = HeunB \left[ \alpha_2 , \beta , \gamma^+ , \delta^+_2 , \,  x_u \right] ;
\nonumber \\*
& &
v_1 = HeunB \left[ \alpha_1 , \beta , \gamma^- , \delta^-_1 , x_v \right]
\; ; \; 
v_2 = HeunB \left[ \alpha_2 , \beta , \gamma^- , \delta^-_2 , x_v \right] ,
\end{eqnarray}
of variables
\[
x_u = - \, \frac{\sqrt{2} M B_*^2 \rho^2}{8 ( \omega+M)} \, , \;
x_v = \frac{\sqrt{2} M B_*^2 \rho^2}{8 ( \omega-M)} \, ,
\]
and parameters
\begin{eqnarray}
& & \alpha_1 = m - \frac{1}{2} \, , \;
\alpha_2 = m + \frac{1}{2} \, , \;
\beta = \sqrt{2} \, , \;
\gamma^{\pm} =
- 2 \left[ \frac{qB_0 ( \omega \pm M)}{M B_*^2} \right]^2 \, , \;
\nonumber \\*
& &
\delta_1^{\pm} = \pm \, \frac{2 \sqrt{2}}{M B_*^2 } ( \omega \pm M ) \left[ \omega^2 - M^2 + \left( m + \frac{1}{2} \right) q B_0 \right] , \, \;
\nonumber \\*
& &
\delta_2^{\pm} = \pm \, \frac{2 \sqrt{2}}{M B_*^2 } ( \omega \pm M ) \left[ \omega^2 - M^2 + \left( m - \frac{1}{2} \right) q B_0 \right]
\end{eqnarray}
and therefore the components of the bi-spinor $\psi$ in (27) are given by
\begin{eqnarray}
& & \psi_1 = \frac{1}{2 \sqrt{\Lambda}} \left \lbrace 1 , \rho^{m- \frac{1}{2}} \right \rbrace u_1 \; , \;
\psi_2 = \frac{1}{2 \sqrt{\Lambda}} \left \lbrace 1 , \rho^{m+ \frac{1}{2}} , \right \rbrace u_2 \; ,
\nonumber \\*
& & \psi_3 = \frac{1}{2 \sqrt{\Lambda}} \left \lbrace 1 , \rho^{m- \frac{1}{2}} \right \rbrace v_1 \; , \;
\psi_4 = \frac{1}{2 \sqrt{\Lambda}} \left \lbrace 1 , \rho^{m+ \frac{1}{2}}
\right \rbrace  v_2 \; .
\end{eqnarray}

Using the expressions (40) in (25), one may compute the radial current density, meaning particles per unit time and per unit covariant 2-surface
\[
d \Sigma_1 = \omega^2 \wedge \omega^3 = \rho d \varphi \wedge d z \, ,
\] 
as
\[
j^1 = i \, \bar{\Psi} \gamma^1 \Psi = \Psi^{\dagger} \alpha^1 \Psi 
= \frac{1}{2 \Lambda} \left[ u_1 v_2 + u_2 v_1 \right]
\]
and the corresponding (radial) current,
\begin{equation}
I ( \rho ) = \int_{-\frac{L_z}{2}}^{\frac{L_z}{2}} \int_0^{2 \pi} e_a^1 j^a \, d \Sigma_1   = \frac{\pi L_z \rho}{\Lambda^2} 
\left[ u_1 v_2 + u_2 v_1 \right] ,
\end{equation}
represented in the figure 1, as a function of
\[
x \approx \frac{\sqrt{2} \, M B_*^2 \rho^2}{8 \omega} \, .
\]
One may notice that, for $x \ll 1$, the current is suddenly increasing from zero to a maximum value, which depends on the ratio $M / \omega$ and on the magnetic field intensity.

\begin{figure}[H]
  \centering
 \includegraphics[width=0.48\textwidth]{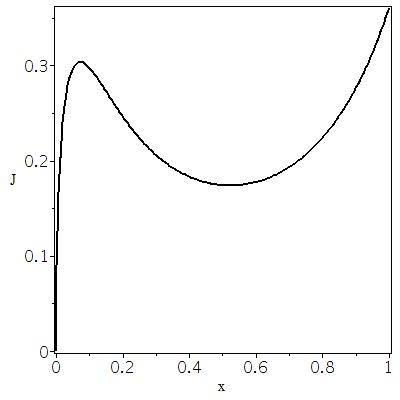} 
  \caption{The radial current (41) in terms of the variable $x \approx \sqrt{2} M B_*^2 \rho^2/(8 \omega)$.}
\end{figure}

The case corresponding to massless fermions is significantly less complicated. Thus, the equation
\begin{equation}
\frac{1}{\Lambda} \left[
\gamma^1 \left( \partial_{\rho} + f \right) + \frac{\Lambda^2}{\rho} \gamma^2 \partial_{\varphi} + \gamma^3 \partial_z + \gamma^4 \partial_t \right]
\Psi_0 - i q \gamma^2 A_2 \Psi_0 = 0 \, ,
\end{equation}
with the variables separation (25),
leads to the following differential equation satisfied by the part depending on $\rho$, i.e.
\begin{equation}
\gamma^1 \left[ \psi_0^{\prime} + f \psi_0 \right] +  i \left \lbrace  \gamma^2 \Lambda
\left[ \frac{m \Lambda}{\rho} - \frac{qB_0 \rho}{2}  \right] +  p \gamma^3 - \omega \gamma^4  \right \rbrace \psi_0 = 0 \, .
\end{equation}

As customary for massless fermions, we are going to use the Weyl representation for the $\gamma$ matrices,
\begin{equation}
\gamma^{\mu} = -i \beta \, \alpha^{\mu} \; , \; \; \gamma^4
= - i \beta \; , \; \; \mu = \overline{1,3}
\, ,
\end{equation}
with
\[
\alpha^{\mu} = \left(
\begin{array}{cc}
\sigma^{\mu} & 0 \\
0 & - \sigma^{\mu} 
\end{array}
\right) ,
\]
and the bi-spinor $\psi_0$ will be taken as
\begin{equation}
\psi_0 ( \rho ) = \left[ \begin{array}{c}
\zeta  ( \rho ) \\ \eta ( \rho )
\end{array} \right] .
\end{equation}
Once the equation (43) decouples in two equations for $\zeta$ and $\eta$,
one gets, for the {\it up} spinor's components, the
second-order differential equations
\begin{eqnarray}
& & \zeta_A^{\prime \prime} + 2 f \zeta_A^{\prime} + \left[ \omega^2 - p^2_z + f^2 -  F^2 
+ \partial_{\rho} \left( f \mp  F \right) \right] \zeta_A = 0 \,   ,
\end{eqnarray}
and similarly for $\eta_A$.
Within the same approximation $B_*^2 \ll qB_0$ and neglecting the powers of $\rho$ larger than 2, 
the equations (46) turn into the simpler forms
\begin{eqnarray}
& & \frac{d^2 \zeta_A}{d \rho^2} + \left[ \frac{1}{\rho} + \frac{B_*^2 \rho}{2} \right] \frac{d \zeta_A}{d \rho} 
\nonumber \\* & &
+ \left[ \omega^2 - p_z^2 + qB_0 \left( m \pm \frac{1}{2} \right)  - \left( m \mp \frac{1}{2} \right)^2 \frac{1}{\rho^2} - \left( \frac{qB_0 \rho}{2} \right)^2 \right] \zeta_A = 0 \, , \nonumber
\end{eqnarray}
whose solutions can be expressed either in terms of confluent hypergeometric functions as [13]
\begin{eqnarray}
\zeta_1 & = & x^{\frac{1}{2} \left( m - \frac{1}{2} \right)} e^{-x/2} U \left[ - \frac{\omega^2 - p_z^2}{2qB_0} + \frac{B_*^2}{4 qB_0} \, , \, m + \frac{1}{2} \, , \, x \right] \, ;
\nonumber \\*
\zeta_2 & = & x^{- \frac{1}{2} \left( m + \frac{1}{2} \right)} e^{-x/2} U \left[ - \frac{\omega^2 - p_z^2}{2qB_0} + \frac{B_*^2}{4 qB_0} - \left( m - \frac{1}{2} \right) \, , \, - \left( m - \frac{1}{2} \right) , \, x \right] \, ,
\nonumber
\end{eqnarray}
or in terms of Whittaker functions [13], as
\begin{equation}
\zeta_1 = \frac{1}{\sqrt{x}}  \, {\rm W}_{\lambda_1 , \mu_1} (x) \; , \;
\zeta_2 = \frac{1}{\sqrt{x}} \, {\rm W}_{\lambda_2 , \mu_2} (x) \; , \;
\end{equation}
with the dimensionless variable
\[
x = \frac{B_*^2 \rho^2}{4} \sqrt{ 1 + \left( \frac{2qB_0}{B_*^2} \right)^2} \approx \frac{qB_0 \rho^2}{2} \, ,
\]
and
parameters
\begin{eqnarray}
& &
\lambda_1 \approx \frac{\omega^2 - p_z^2}{2qB_0} - \frac{B_*^2}{4 qB_0} + \frac{1}{2} \left( m + \frac{1}{2} \right) , \; \; \mu_1 = 
\frac{1}{2} \left( m- \frac{1}{2} \right) ;
\nonumber \\*
& &
\lambda_2 \approx \frac{\omega^2 - p_z^2}{2qB_0} - \frac{B_*^2}{4 qB_0} + \frac{1}{2} \left( m - \frac{1}{2} \right) , \; \; \mu_1 = 
\frac{1}{2} \left( m  + \frac{1}{2} \right) . \nonumber
\end{eqnarray}

\section{Conclusions}

Within the framework of the gauge-invariant geometry, based on the semi-direct product of the local groups $SO(3,1)$ and $U(1)$, the present paper is focusing on the Klein--Gordon and Dirac equations describing particles evolving in a background endowed with the Melvin's metric.

By making use of the Cartan's formalism, we have derived the corresponding Einstein--Melvin equations leading to the essential relation between the model's parameters (14).
As a remark, switching to the canonical bases, the third covariant induction component is given by the expression
\[
B_z = \sqrt{|g|} F_{(c)}^{12} =  \sqrt{|g|} \, e_a^1 e_b^2 F^{ab} \equiv B_0 \, .
\] 

In case of bosons, the Klein--Gordon equation can be integrated exactly, its solution being given by the Heun biconfluent functions of parameters (20).

The equations (33) and (34) coming from the Dirac equation (24) have rather complicated expressions, containing several additional terms which were neglected in [6].
In the assumption $B_*^2 \ll q B_0$, we have been able to find solutions for the equations (35) and (36), expressed in terms of Heun's biconfluent functions (38). 
The corresponding radial particle-current (41), represented in the figure 1, is starting from zero, at the origin $\rho =0$ and exhibits a rather non-trivial behavior, characterized by a sudden growth to a maximum value. This one is followed by a local minimum which might signal, in the approximation we have used, the presence of a plateau. The seemingly far away unlimited increasing is a result of the violation of the approximation holding condition which demands $x < 1$.

\begin{flushleft}
\begin{Large}
{\bf Acknowledgement}
\end{Large}
\end{flushleft}
This work was supported by a grant of Ministery of Research and Innovation, CNCS - UEFISCDI, project number PN-III-P4-ID-PCE-2016-0131, within PNCDI III.
The authors declare that there is no conflict of interest regarding the publication of this paper.

\end{document}